\documentclass[aps,showpacs,a4paper,floatfix,twocolumn,prc,amsmath,amssymb,superscriptaddress]{revtex4-1}
\usepackage{graphicx}
\usepackage{epsfig}
\usepackage{ulem}
\usepackage{morefloats}
\usepackage{rotating}
\usepackage{xcolor}
\usepackage{soul}

\begin{document}

\title{Clustered Nature of Hot and Dense Nuclear Matter: A quantum statistical approach}
\author{G. R\"opke}
\email{gerd.roepke@uni-rostock.de}
\affiliation{Institut f\"ur Physik, Universit\"at Rostock, Albert-Einstein-Strasse 23-24, 18059 Rostock, Germany.}
\author{H. Pais}  
\email{hpais@uc.pt}
\affiliation{CFisUC, Department of Physics, University of Coimbra, 3004-516 Coimbra, Portugal.}
\author{J. B. Natowitz}
\email{natowitz@comp.tamu.edu}
\affiliation{Cyclotron Institute, Texas A\&M University, College Station, Texas 77843, USA; Ret.} 
\author{D. Blaschke}
\email{david.blaschke@uwr.edu.pl}
\affiliation{Institute of Theoretical Physics, University of Wroclaw, Max Born place 9, 50-204 Wroclaw, Poland}
\affiliation{Center for Advanced Systems Understanding (CASUS), Untermarkt 20, 02826 Görlitz, Germany}
\affiliation{Helmholtz-Zentrum Dresden-Rossendorf (HZDR), Bautzner Landstrasse 400, 01328 Dresden, Germany}
\begin{abstract}
The equilibrium abundances of the light clusters $^2$H, $^3$H, $^3$He, $^4$He in hot nuclear matter at densities near the saturation density are of essential interest for nuclear physics and astrophysical applications, but theoretical approaches give diverging answers. We compare the quantum statistical approach with the recently discussed phase-space excluded-volume approach. We analyze the main ingredients, the Mott momentum, and the momentum distribution functions of light clusters. We observe a sharp decrease in cluster abundances as the density approaches saturation density, that is also seen in a relativistic mean-field calculation.
We outline possible improvements in determining the composition of hot, dense matter in thermodynamic equilibrium. Non-equilibrium effects must be taken into account to investigate cluster formation in heavy-ion collisions.
%To investigate cluster formation in heavy-ion collisions, non-equilibrium effects must be taken into account.
\end{abstract}

\maketitle

\section{Introduction}

The composition of hot and dense matter, in particular the mass fraction of $\alpha$ particles as the most bound few-particle state, was considered in 
Ref. \cite{Blaschke25}, based on a quantum statistical (QS) approach which includes single-particle self-energy shifts and Pauli blocking for the few-nucleon states.
As a consequence, the properties of the bound states are modified by the in-medium effects. 
The binding energies of the bound states are weakened by the Pauli blocking mechanism, and with increasing density, finally they merge with the continuum. This is denoted as the Mott effect \cite{R82}. However, beyond this critical density, correlations remain in the continuum and behave like a resonance state. 

In this paper we consider hot and dense matter in the range of saturation density (baryon number density $n_B \approx 0.16$ fm$^{-3}$) and temperatures up to 50 MeV,
where correlations between the constituents such as cluster formation are relevant. %Nucleons form clusters. Pions and further particles appear at increasing density of energy.
%Cluster formation is 
The formation of bound states of nucleons is of interest for nuclear structure and reactions, for heavy ion collisions, and also for various astrophysical phenomena.
In particular, we are interested in the formation of %bound states, the 
light clusters $\nu$: deuteron ($^2$H, $d$), the triton ($^3$H, $t$), the helion ($^3$He, $h$), and the $\alpha$ particle.
Few-nucleon correlations determine not only the equations of state and all related thermodynamic properties, but also transport properties and kinetic processes.
We do not review the multiple work on the cluster formation in hot and dense matter in this paper, but focus on some recent work.

\begin{figure*}[ht]
\begin{minipage}[b]{0.5\linewidth}
\centering
\includegraphics[width=1.2\textwidth]{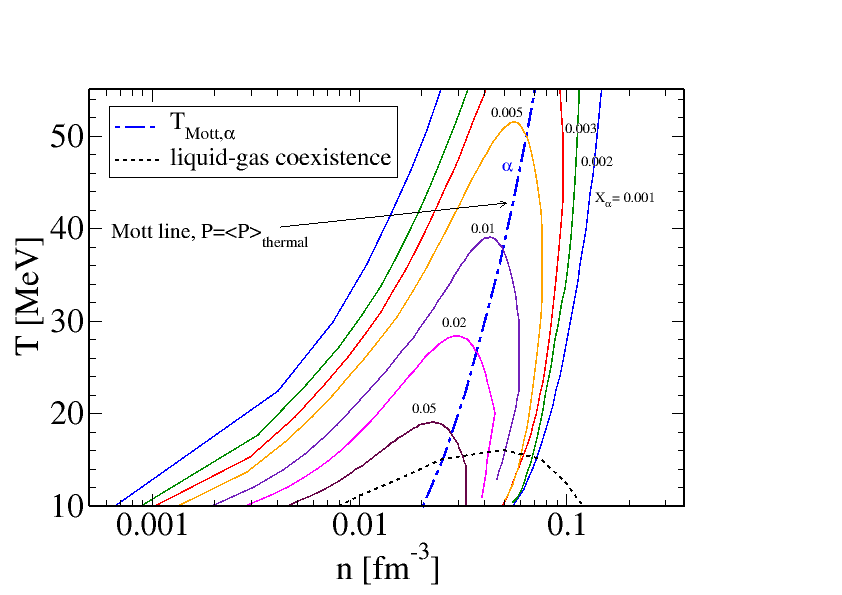}
\caption{Contour plots for $X_\alpha$ in the $(n_B,T)$ plane for different values of the mass fraction of the $^4$He cluster for symmetric nuclear matter, adapted from \cite{Blaschke25}.}
\label{fig:PLB25}
\end{minipage}
\hspace{0.5cm}
\begin{minipage}[b]{0.45\linewidth}
\centering
\includegraphics[width=\textwidth]{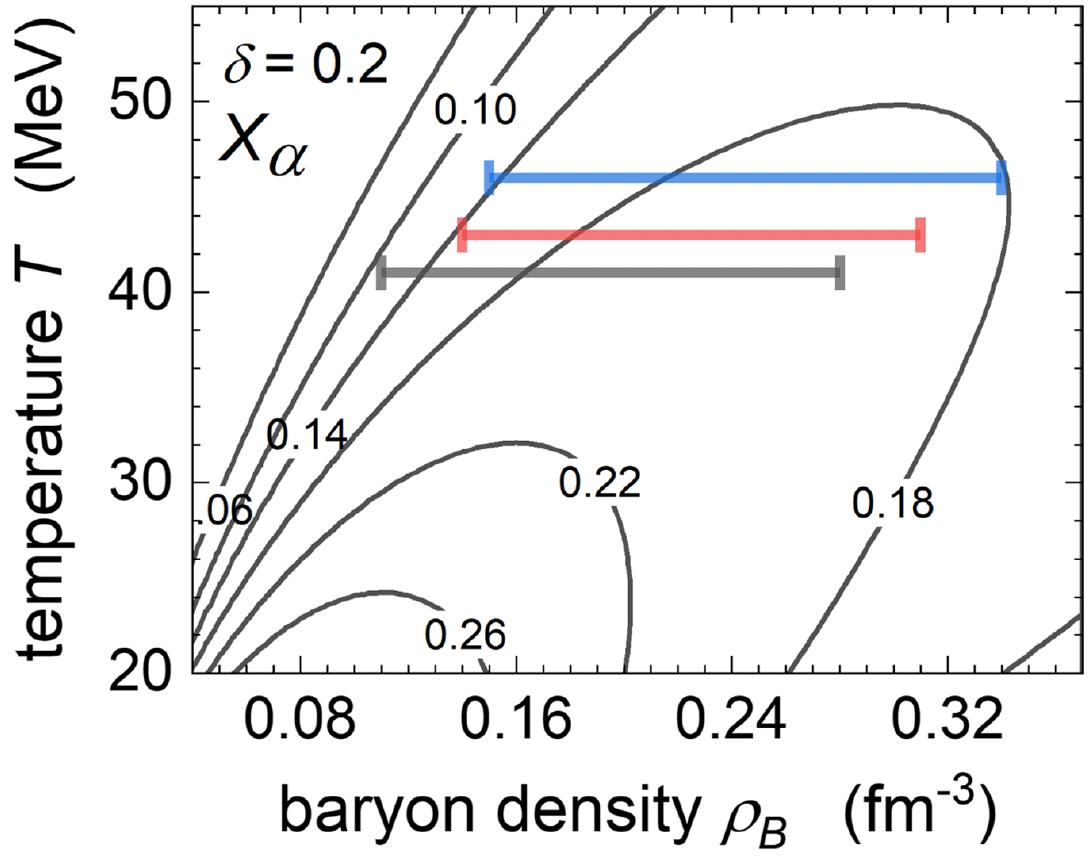}
\caption{Contour plots for $X_\alpha$ in the $(n_B,T)$ plane for different values of the mass fraction of the $^4$He cluster for $\delta=0.2$, from \cite{Wang26}.}
\label{fig:PRL26}
\end{minipage}
\end{figure*}

Within the quantum statistical approach, the spectral function can be introduced, and for the density a generalized
Beth-Uhlenbeck formula can be derived \cite{Schmidt90,Blaschke26}, whereby few-particle states are obtained from the solution of an in-medium Schr\"odinger equation. 
Shifts for the light clusters are calculated as function of the thermodynamic parameters temperature $T$, baryon number density $n_B$, proton fraction $Y_p$, as well as the total momentum $\bf P$  \cite{Roepke11}.
These shifts define the Mott point $n_\nu^{\rm Mott}(T,Y_p)$, which is, for given $T,Y_p$, the density at which for  $P=0$ the bound state merges with the continuum \cite{Hagel12}.
Above this critical density, the finite value $P^{\rm Mott}_\nu(T,n_B,Y_p)$ can be defined so that clusters with $P>P^{\rm Mott}_\nu$ remain bound. 
The composition of symmetric nuclear matter at $T=5, 10, 15, 20$ MeV as a function of the baryon density, calculated from the quantum statistical approach, is shown in \cite{Roepke15} where also the continuum contributions are taken into account.

Since the quantum statistical approach is very complex, simpler approximations have been used to describe the in-medium effects.
Self-energy shifts are parametrized by the Skyrme interaction or relativistic mean-field (RMF) approximation.  
The generalized RMF approach includes also the light clusters, considered as new quasiparticles with energies which are shifted by medium effects \cite{Typel10,PaisPRL}.

The effect of Pauli blocking is often approximated by the concept of excluded volume \cite{Hempel10}. 
However, Pauli blocking is strongly dependent on the center-of-mass momentum $\bf P$ of the bound state so that the dissolution of the bound state depends not only  on $n_B, T, Y_p$, but also on $\bf P$. 
This leads to the concept of the phase-space excluded volume (PSEV) approach used in  Ref. \cite{Wang2026}.
This concept has been applied to discuss light particle production from intermediate energy heavy-ion collisions in a recent letter \cite{Wang26}.
There, the formulation of the PSEV approach leads to different results than those obtained with the QS approach. See the contour plots of the mass fractions $X_\alpha$ of $\alpha$ particles presented in Figs. \ref{fig:PLB25} and \ref{fig:PRL26}.
Since both QS  and PSEV approaches have been used to explore the composition of hot and dense nuclear matter
%, and comparison should be performed. 
and their differing results have important implications for both nuclear physics and various astrophysical phenomena, %\cite{Wang26},
a more detailed discussion is warranted.

\section{The quantum statistical approach}

Within a Greens function approach, the spectral function is introduced, and a cluster decomposition of the self-energy leads to the in-medium Schr\"odinger equation \cite{R82}. We adopt the notation of \cite{Wang26},
\begin{eqnarray}
\label{mediumSEq}
 &&   \left[\sum_i^A\epsilon_{\tau_i}({\bf p}_i)-\epsilon_\nu({\bf P})\right] \Psi_{\nu,{\bf P}}(1,\dots ,A)\nonumber\\
 &&+\sum_{i<j}\left[1-f^{\rm tot}_{\tau_i}({\bf p}_i)-f^{\rm tot}_{\tau_j}({\bf p}_j)\right] \int \prod_l^{1',... ,A'} \frac{d^3p_l}{(2 \pi \hbar)^3} \nonumber \\
 && \times V(ij,i'j') \prod_{k \neq i,j} \delta_{kk'}\Psi_{\nu,{\bf P}}(1',\dots ,A')=0,
\end{eqnarray}
where $\epsilon_{\tau}({\bf p})$ is the nucleon quasiparticle energy ($\tau=n,p$), the Pauli blocking term 
$f^{\rm tot}_{\tau}$ represents the total occupation of phase space by nucleons in the nuclear medium,
and $\epsilon_\nu({\bf P}), \Psi_{\nu,{\bf P}}(1,\dots ,A)$ are the eigenvalue and eigenfunction of the $A$-particle cluster.
With the spectral function, the equation of state for the nucleon number densities $n_\tau$ as a function of $T,\mu_n,\mu_p$ is derived, and gives the Beth-Uhlenbeck formula for the second virial coefficient, see \cite{SM}. According to the generalized Beth-Uhlenbeck formula \cite{Schmidt90}, the total nucleon density $n_B(T,\mu_\tau)$ includes the contribution from single-nucleon quasiparticles. 
In addition, there is a correlation term for density that can be decomposed into a contribution from bound states and a contribution from continuum correlations (expressed, e.g., by in-medium phase shifts). 
This decomposition is not unique; using the Levinson theorem, part of the contribution from continuum correlations can be transferred to the contribution from bound states. 

The treatment of continuum correlations presents a problem in all approximations. 
Whereas well-defined resonances such as $^5$He can be treated as a new component of the nuclear system, other correlations shown by the phase shifts are often omitted. Within the framework of a quantum statistical approach, the total amount of a specific channel characterized by the number of neutrons, protons, total momentum and angular momentum contains the bound part and the contribution of the continuum, as known from the generalized Beth-Uhlenbeck formula.
 Approximations such as gRMF theory can simulate the contribution of the continuum as a shift of the corresponding few-particle state.
The PSEV approach of \cite{Wang26} does not provide a general approach for treating continuum correlations. 
However, going to high temperatures up to 50 MeV, the contribution of the continuum becomes more significant.
%Like the in-medium shift of the energies of light clusters, the contribution of the continuum may be included phenomenologically in the PSEV approach \cite{Wang26} in future work.

To compare the differing approaches employed to extract EOS information we focus on the  mass fraction of $\alpha$-particles as the most strongly bound light clusters. 
Calculations based on the quantum statistical approach show a dissolution of $\alpha$-particles at lower temperatures at about $n_B \approx 0.03$ fm$^{-3}$ \cite{THSR}. With increasing temperature $\alpha$-particles can be found also at higher densities, but are strongly reduced when the baryon density approaches the saturation  density, see \cite{Blaschke25} and Fig. \ref{fig:PLB25}. (In addition to the contour plot of constant $\alpha$-mass fractions $X_\alpha$, further lines are shown in Fig. \ref{fig:PLB25} which are explained in \cite{Blaschke25}.)

Reference \cite{Wang26} addressed the topic of cluster formation in hot, dense nuclear matter, with a particular focus on the Mott effect.
In Fig. 2 of their letter, also shown in this work (see Fig.~\ref{fig:PRL26} above), the authors presented unexpectedly abundant $\alpha$-clustering, 
with $\alpha$ mass fraction $X_\alpha$ in the range 0.18 - 0.26 in hot nuclear matter at densities around 1 to 2 times the nuclear saturation density, for $20<T<40$ MeV.  
(Their calculations give only a weak dependence on the isospin asymmetry $\delta = 1-2Y_p$, the value $\delta=0.2$ was adapted to Au + Au collisions.)
In Ref. \cite{Blaschke25},   $\alpha$-particles appear significantly only below the saturation density, and their abundances are lower than those reported in Ref. \cite{Wang26}, see Fig.~\ref{fig:PLB25}.

\section{The Mott momentum}
To understand the different results for the composition of hot and dense matter, we investigate the key ingredients of the various approaches. This can help to find improved approaches.
A crucial point in \cite{Wang26} is the determination of the Mott momentum $P^{\rm Mott}_\nu(T,n_B,Y_p)$. 

The in-medium few-nucleon Schr\"odinger equation (\ref{mediumSEq}) determines the shifts of the bound state energies $\epsilon_\nu({\bf P};T,n_B,Y_p)$, the excited states, the scattering phase shifts, and the bound state wave functions ($\nu = d,t,h,\alpha$).
In particular, depending on $\bf P$, the bound state merges at increasing density with the continuum of scattering states and disappears.
Assuming the dissolution into the constituents (we assume that this is the decay channel with the lowest energy), this condition is given by 
\begin{eqnarray}
\label{Mott}
    \epsilon_\nu(\hat P;\hat T,\hat n_B,\hat Y_p)&=&N\epsilon_n(\hat P/A;\hat T,\hat n_B,\hat Y_p)\nonumber \\ &&+Z\epsilon_p(\hat P/A;\hat T,\hat n_B,\hat Y_p),
\end{eqnarray}
which we denote as Mott condition.
Since the self-energies of the nucleons also appear in the in-medium Schr\"odinger equation,
this condition concerns the contribution of the Pauli blocking (the contribution of the effective mass was considered in \cite{Roepke09}).
With increasing density, the Mott condition occurs first at $P=0$ at the Mott density $n_{\nu}^{\rm Mott}(T,Y_p)$ since the overlap of the bound state wave function with the Fermi distribution is a maximum.
At higher densities $n_B>n_{\nu}^{\rm Mott}$, we can introduce a Mott momentum $P^{\rm Mott}_\nu(T,n_B,Y_p)$ which fulfills the condition (\ref{Mott}).
Bound states of clusters $\nu$ can exist only for $P > P^{\rm Mott}_\nu$.
We discuss different approaches to determine the Mott momentum.

\begin{figure}[t]
  \centering
  \includegraphics[width=0.5\textwidth]{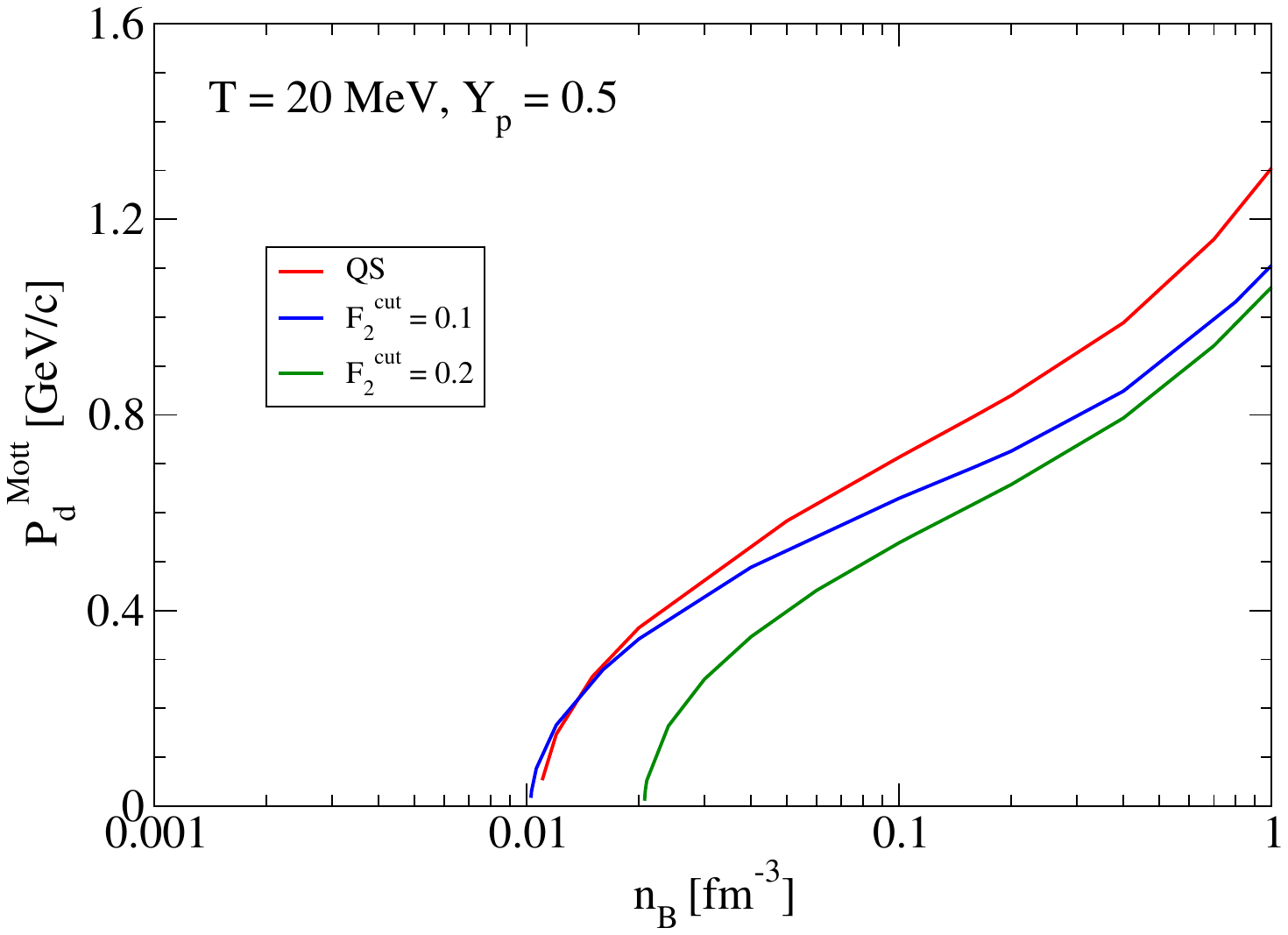}
\caption{Mott momentum $P^{\rm Mott}_d$ for deuterons in symmetric matter at $T=20$ MeV. 
The quantum statistical approach (red) is compared with the phase-space excluded-volume approach for two values of $F^{\rm cut}_2$.}
\label{fig:PMottd}
\end{figure}

\begin{figure}[t]
  \centering
  \includegraphics[width=0.5\textwidth]{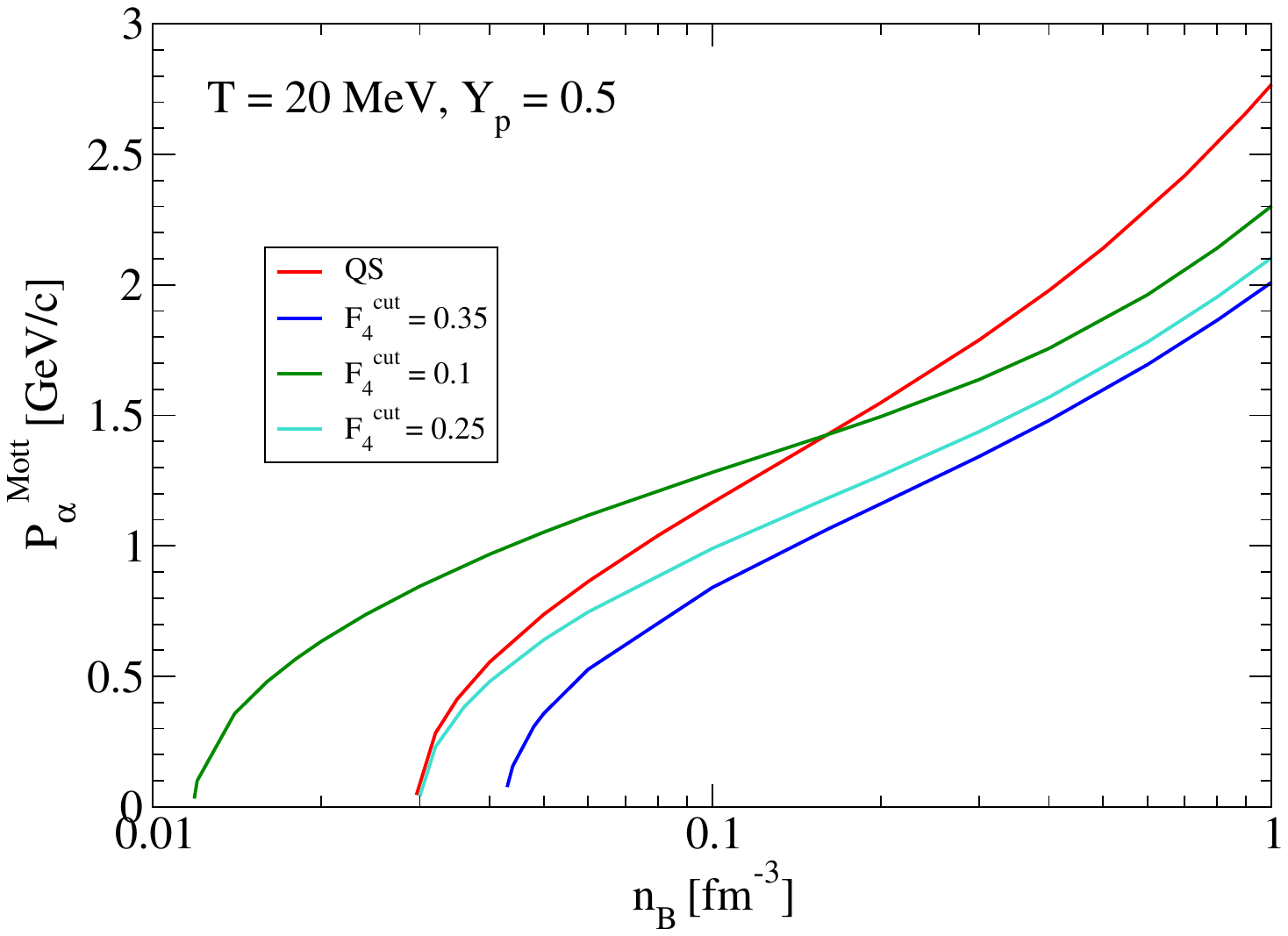}
\caption{Mott momentum $P^{\rm Mott}_\alpha$ for $\alpha$ particles in symmetric matter at $T=20$ MeV. 
The quantum statistical approach (red) is compared with the phase-space excluded-volume approach for three values of $F^{\rm cut}_4$.}
\label{fig:PMotta}
\end{figure}

\subsection{The phase-space excluded-volume approach}

In contrast to the excluded volume concept which considers Pauli blocking in a very crude approximation, the phase-space excluded-volume approach of Ref. \cite{Wang2026} considers Pauli blocking in phase space. 
The overlap integral of the bound state wave function  with occupation numbers of the surrounding matter is introduced as
\begin{equation}
\label{overlap}
\langle f_\tau \rangle_\nu({\bf P})= \int f^{\rm tot}_\tau({\bf p}) |\phi_{\nu, {\bf P}}({\bf p})|^2\frac{d^3 {\bf p}}{(2 \pi \hbar)^3},
\end{equation}
where $\tau= n$ or $p$, and $|\phi_{\nu, {\bf P}}({\bf p})|^2$ denotes the normalized one-body probability distribution of the nucleons inside the light cluster $\nu$. This overlap integral measures the relative amount of the doubly occupied states in momentum space and thus the violation of the Pauli principle.
The use of the free nucleon wave function $\phi_{\nu, {\bf P}}({\bf p})$ and the expressions for the occupation numbers $f^{\rm tot}_\tau({\bf p})$ are approximations which can be improved. 
An important diagnostic is the dependence of the overlap integral on the total momentum $\bf P$ of the cluster.

The phase-space excluded-volume approach was introduced in Ref. \cite{Wang2026} to reduce computational effort in solving kinetic equations, considering the overlap integral (\ref{overlap}) as a relevant parameter.
The Mott condition is approximated by the condition 
\begin{equation}
\label{PMottex}
   \langle f_\tau \rangle_\nu({\bf P_\nu^{\rm Mott, cut}})= F_A^{\rm cut}.
\end{equation}
The cutoff value $F_A^{\rm cut}$ is an empirical parameter which depends on the the mass number $A$ of the nucleus.

The use of the overlap integral of the bound state wave function and the phase space occupation of the nuclear medium, Eqs. (\ref{overlap}) and (\ref{PMottex}), to determine the Mott momentum $P^{\rm Mott}_\nu$ is motivated by the Pauli principle, see Fig. 1 of \cite{Ropke20}.
It has the advantage that it considers the dependence of the Pauli blocking shift $\Delta E_\nu^{\rm Pauli}(P)$ on the total momentum of the cluster. 
Furthermore, it is not dependent on the choice of an interaction potential, but depends only on measurable quantities, the bound state wave function and the occupation numbers in phase space.
However, there is no derivation of this simplifying relation from basic quantum statistics.

To determine the empirical parameter $F_A^{\rm cut}$, two possibilities can be considered: 
(i) The comparison of $\bf P_\nu^{\rm Mott, cut}$ according eq. (\ref{PMottex}) with $\bf P_\nu^{\rm Mott}$ of the solution of the in-medium Schr\"odinger equation, 
and (ii) the fit to data from HIC experiments, as proposed in \cite{Wang26} with the proposed values ${\bf F}^{\rm cut}=(0.192, 0.248, 0.345)$ for $A=(2,3,4)$. 
Both choices have consequences for the equilibrium properties of nuclear matter, as pointed out in this work.

The comparison with the in-medium Schr\"odinger equation has been performed in \cite{Wang2026} for the deuterons.
A rough overall agreement of ${\bf P}_d^{\rm Mott, cut}$ with ${\bf P}_d^{\rm Mott}$ was reported for $F_2^{\rm cut}$ in the range between 0.15 and 0.2.
We repeat the calculations in Fig. \ref{fig:PMottd} and compare with the expression for the Pauli blocking shifts according to \cite{Roepke11}. 
For the few-particle wave function we used Jacobian coordinates to separate the c.m. momentum $\bf P_\nu$.
Results for $\bf P_\nu^{\rm Mott, cut}$ of symmetric matter at $T=20$ MeV are shown as a function of the density $n_B$ for different values of $F_A^{\rm cut}$.
The Mott densities $n_d^{\rm Mott}$ of both approaches coincide for $F_2^{\rm cut} \approx 0.1$.
For a detailed comparison of the Mott momenta obtained from both the quantum statistical and the phase-space excluded-volume approach see \cite{SM}.% Appendix \ref{app:1}.

The same calculation was performed for the $\alpha$ particles; see Fig. \ref{fig:PMotta}.
A significant difference is observed for the cutoff parameter $F_4^{\rm cut}=0.35$.
Compared to the QS approach, the region in which bound states can exist extends to higher densities.

\subsection{An exact relation for the deuteron Mott momentum}

An exact expression can be given for the deuteron case in perturbation theory \cite{Roepke09}.
Starting from the free deuteron case, after separation of the c.m. motion by using Jacobian coordinates, we have with the relative momentum ${\bf q}=({\bf p}_2-{\bf p}_1)/2$
\begin{equation}
\label{dSgl}
\left(\frac{q^2}{m}-E_d^0\right)\psi^0_d({\bf q})+\int \frac{d^3{\bf q'}}{(2 \pi \hbar)^3} V({\bf q,q'}) \psi^0_d({\bf q}')=0    
\end{equation}
with the nucleon mass $m$ and the bound state energy $E_d^0=-2.225$ MeV.
In perturbation theory, the Pauli blocking shift reads
\begin{eqnarray} 
\label{EPauld}
    &&\Delta E_d^{\rm Pauli}(P)=-\int \frac{d^3{\bf q}}{(2 \pi \hbar)^3}\int \frac{d^3{\bf q'}}{(2 \pi \hbar)^3}\psi^{0,*}_d({\bf q})\nonumber \\
    &&\times \left[f_n\left(\frac{{\bf P}}{2}+{\bf q}\right)+f_p\left(\frac{{\bf P}}{2}-{\bf q}\right)\right]V({\bf q,q'}) \psi^0_d({\bf q}')\nonumber \\
    && =\int \frac{d^3{\bf q}}{(2 \pi \hbar)^3} \left[f_n\left(\frac{{\bf P}}{2}+{\bf q}\right)+f_p\left(\frac{{\bf P}}{2}-{\bf q}\right)\right]\nonumber \\
    &&\times\left(\frac{q^2}{m}-E_d^0\right)|\psi^0_d({\bf q})|^2
\end{eqnarray}
where the Schr\"odinger equation (\ref{dSgl}) was used.
The Mott momentum follows from the condition 
\begin{equation}
\label{pmottqsd}
    \Delta E_d^{\rm Pauli}(P^{\rm Mott}_d)=-E^0_d
\end{equation}
for the density above the Mott density.
Note that this condition also depends only on the deuteron wave function and the occupation numbers in phase space which are measurable quantities.
This is also the case for Eq. (\ref{PMottex}), but in contrast, no open parameter $F^{\rm cut}_2$ appears.
The occupation numbers $f_\tau(p)$ can be equilibrium distributions for arbitrary $T,n_B,Y_p$, but also non-equilibrium distributions.

\begin{figure}[t]
  \centering
  \includegraphics[width=0.5\textwidth]{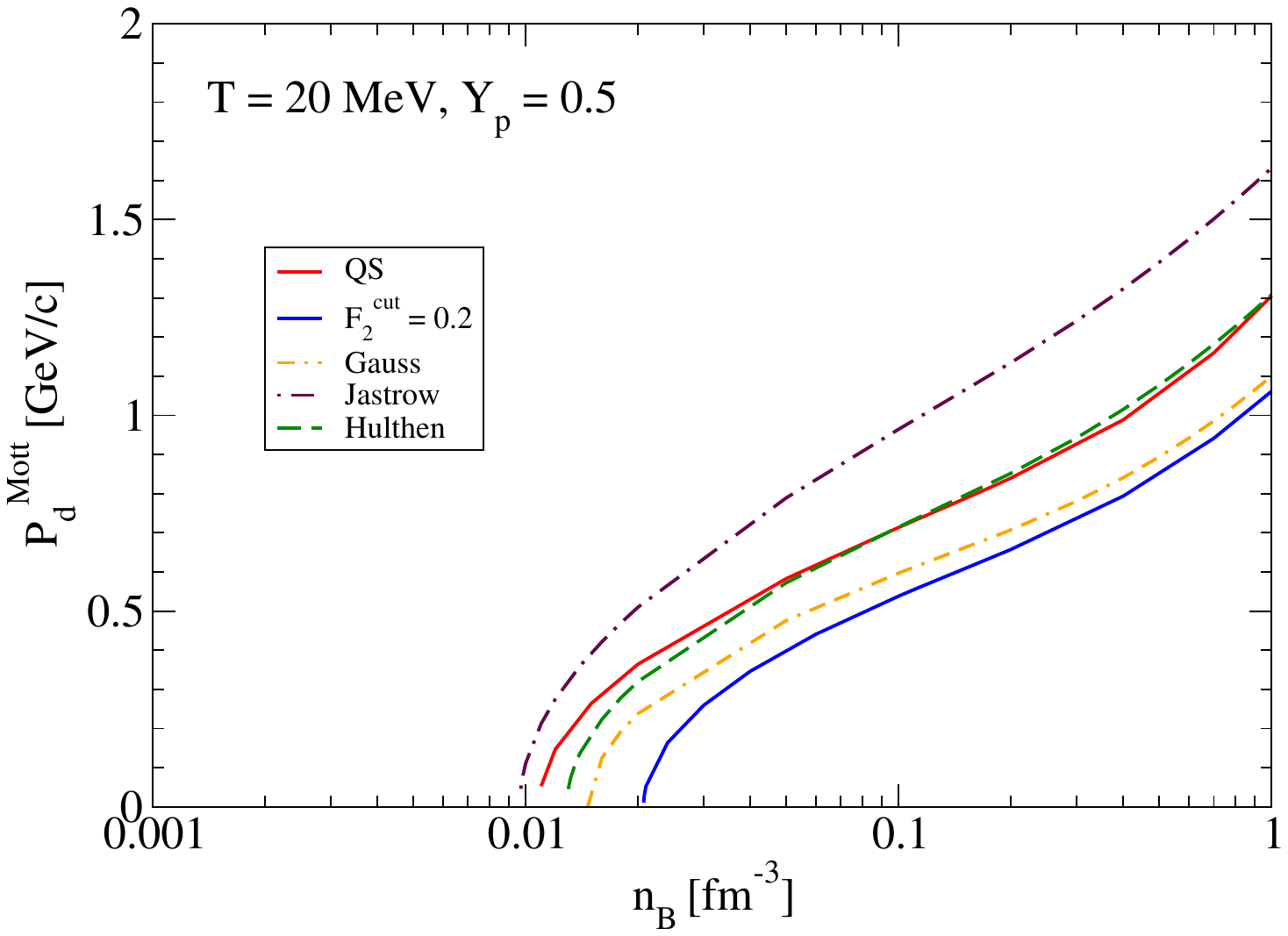}
\caption{Mott momentum $P^{\rm Mott}_d$ for deuterons in symmetric matter at $T=20$ MeV. 
The quantum statistical approach (red) is compared with the Gauss and Jastrow wave functions.}
\label{fig:PMottdJ}
\end{figure}
\iffalse
\begin{figure}[t]
  \centering
  \includegraphics[width=0.5\textwidth]{pmotta.pdf}
\caption{Mott momentum $P^{\rm Mott}_\alpha$ for $\alpha$ particles in symmetric matter at $T=20$ MeV. 
The quantum statistical approach (red) is compared with the phase-space excluded-volume approach for three values of $F^{\rm cut}_4$.}
\label{fig:PMotta}
\end{figure}
\fi 

For the evaluation of Eq. (\ref{EPauld}), the deuteron wave function $\psi^0_d({\bf q})$ is required.
A simple approximation is a Gaussian form, adapted to the point rms radius of the deuteron, see \cite{SM}. % appendix \ref{app:1}.
Good agreement with the phase-space excluded-volume calculations for $F_2^{\rm cut}=0.1$ is found.
However, this approximation of the deuteron wave function by a Gaussian form is not very good, and more sophisticated approaches such as the Jastrow form \cite{Roepke09} give a different result, 
see \cite{SM} % appendix \ref{app:1} 
and Fig. \ref{fig:PMottdJ}.
Another often used approximation for the deuteron wave function is  the Hulthen form.
Calculations for the Hulthen wave function $u(r) =N(e^{-\gamma r}-e^{-\beta r})$, $\gamma =0.2316$ fm$^{-1}$, $\beta = 5.98 \gamma$ are also shown in \cite{SM} % appendix \ref{app:1} 
and Fig. \ref{fig:PMottdJ}.
The Jastrow wave function, similar to the Hulthen wave function for the deuteron, is more extended in momentum space so that the overlap with the Fermi distribution is less dependent on $P$.
The QS result goes beyond the lowest order of perturbation theory so that the Mott momentum comes out as shown in Fig. \ref{fig:PMottd}.
Our result is as follows:\\
(i) We don't need a cutoff parameter $F^{\rm cut}_A$. The overlap equation (\ref{overlap}) of the bound state wave function with the distribution in phase space should be replaced by another expression (\ref{pmottqsd}).\\
(ii) The form of the deuteron wave function in momentum space is essential.
A Gaussian form is not very accurate to calculate the Mott momentum.\\

Instead of Eq. (\ref{overlap}), we propose using the parameterization of the Mott momentum according to Ref. \cite{Roepke11} or according to Eqs. (\ref{EPauld}), (\ref{pmottqsd}) with a suitable expression for the deuteron wave function.
Equation (\ref{overlap}) can be regarded as a certain approximation that describes the part of Equation (\ref{EPauld}) containing $E_d^0$.
One choice for determining the cutoff parameter $F_2^{\rm cut}$ could be to adjust it to the QS value for the Mott density $n_d^{\rm Mott}$; see Fig. \ref{fig:PMottd}.

\subsection{Expression for $A=3,4$}

The treatment of the other light nuclei $t, h, \alpha$ is more complex. We can also separate the motion of the center of mass by introducing Jacobi coordinates, but eliminating the interaction potential, as was shown for the deuteron case, is generally not possible.
Expressions for the momentum dependence of the Pauli blocking shift are given for the QS approach in \cite{Roepke15}.
The bound state vanishes when the Pauli blocking shift is equal to the binding energy.
For $n_B > n_B^{\rm Mott}$ at given $T, Y_p$, the corresponding Mott momentum for the $\alpha$ particles is shown in Fig. \ref{fig:PMotta}.

As in the case of deuterons,  for a given momentum $P$, the quantum statistical approach \cite{Roepke15} leads to the $\alpha$-particles being resolved at lower densities  than under the phase-space excluded-volume approach  ($F_4^{\rm cut}=0.35$).
Good agreement with the QS value for the Mott density $n_\alpha^{\rm Mott}$ is observed at $F_4^{\rm cut}=0.25$.
Reducing the parameter space in which $\alpha$-particles can exist leads to lower abundance values.
The values for the cutoff parameters $F_A^{\rm cut}$ proposed in Refs.~\cite{Wang26,Wang2026} expand the region in phase space where bound states can exist compared to the QS approach.
This is one reason for the discrepancies in the abundances of light nuclei; see Figs. \ref{fig:PLB25} and \ref{fig:PRL26}.\\

\begin{figure*}[t]
\begin{minipage}[b]{0.5\linewidth}
\centering
\includegraphics[width=\textwidth]{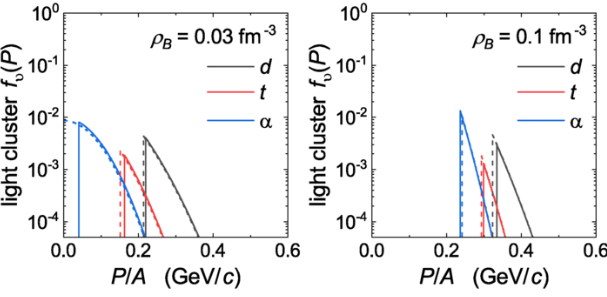}
\caption{Momentum distribution functions for deuterons, tritons, and $\alpha$ particles,
taken from Ref.~\cite{Wang2026}, Fig. 8 there.
Symmetric nuclear matter is considered at two baryon density $n_B =  0.03$ and 0.1 fm$^{-3}$ and temperature $T = 20$ MeV.
The scale of $f_\nu(P)$ is reduced to $f_\nu > 10^{-4}$.}
\label{fig:dist-funct-wang}
\end{minipage}
\hspace{0.5cm}
\begin{minipage}[b]{0.45\linewidth}
\centering
\includegraphics[width=\textwidth]{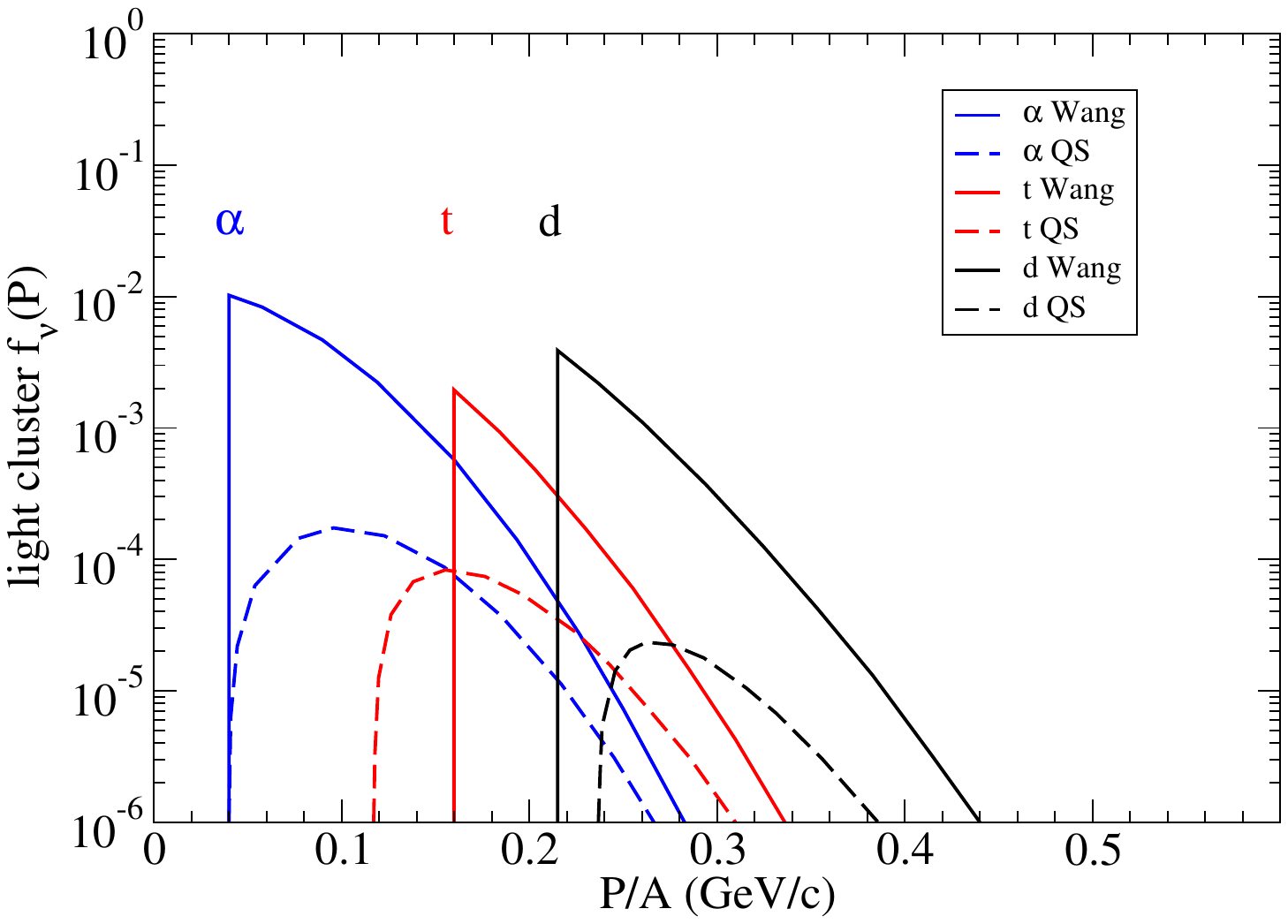}
\caption{Distribution functions of the light clusters.
The calculations using the approach of Wang et al.~\cite{Wang2026} omitting the Pauli-blocking shift of the in-medium binding energies is compared with the quantum statistical (QS) calculation which accounts for the Pauli blocking shifts \cite{Roepke15}.}
\label{fig:distQS}
\end{minipage}
\end{figure*}

As discussed for the deuteron case, the use of an appropriate $\alpha$-particle wave function gives a different result for the Mott momentum. 
Whereas the Gaussian approximation gives a larger range for the existence of bound states, this range is reduced for more realistic wave functions, see Fig. \ref{fig:PMotta}.
In particular, the use of Eq. (\ref{overlap}) with cutoff-parameter $F_4^{\rm cut}=0.35$ allows for $\alpha$ bound states in a region where they cannot exist according to the QS approach.
The numerical solution of the in-medium Schr\"odinger equation (\ref{mediumSEq}) would give the correct values of the Mott momentum $P^{\rm Mott}_\nu$.

\section{The light-cluster distribution functions}

Another important aspect is the determination of the momentum distribution function.
The quantum statistical approach employs the generalized Beth-Uhlenbeck formula \cite{Schmidt90,Blaschke26}, 
in which the contribution of various channels to the total baryon density is specified within the framework of the quasiparticle approach, 
taking into account excited states and continuum correlations.
The quasiparticle energies of the nucleons can be approximated using Skyrme or RMF expressions.
For bound states, medium-modified quasiparticle energies can be introduced, which include Pauli blocking shifts in addition to the nucleon self-energies.
Continuum contributions are related to scattering phase shifts. The contributions from the various channels are complex; an estimate can be found in \cite{Roepke15}.

In the framework of the phase-space excluded-volume approach \cite{Wang26,Wang2026}, the following expression for the distribution of light clusters is given:
\begin{equation}
\label{distribution}
    f^{\rm eq}_\nu({\bf P})=\frac{H(|{\bf P}|-P^{\rm Mott}_\nu)}{\exp\left[\frac{\epsilon_\nu ({\bf P})-\mu_\nu}{k_BT}\right]\pm 1}
\end{equation}
with the Heaviside step function $H$. The energies $\epsilon_\nu$ result from an extended Skyrme interaction, and the corresponding shift can be transferred to the chemical potentials $\mu_\nu$.
Examples of the momentum distribution functions for light clusters are shown in Fig.~\ref{fig:dist-funct-wang}, taken from Ref.~\cite{Wang2026},
a similar picture for symmetric matter at a baryon density $n_B=0.16$ fm$^{-3}$ and a temperature $T=20$ MeV is shown in Fig. 2 of the Supplementary material for Ref.~\cite{Wang26}.
As a function of $P$, a sharp jump from zero is obtained at $P^{\rm Mott}$, and for larger values the distribution function agrees with that of free clusters.
However, this jump behavior is an artifact; the distribution function approaches zero gradually when $P$ falls below the Mott momentum.

While the explicit consideration of the Mott momentum represents an advancement of the phase-space excluded-volume approach compared with other empirical approaches that deal with light clusters in dense matter, further improvements are required.
Instead of the quasi-particle energies used in (\ref{distribution}), which take only the Skyrme interaction into account, medium-modified quasi-particle energies $\epsilon_\nu(P)$ that account for Pauli blocking should be used.
As pointed out in Ref. [6], p.4, the binding energy of
the light clusters that enters the energy $\epsilon_\nu(P)$ is taken
as its value in free space.
This contradicts the disappearance of the binding energy at the Mott momentum.

A more fundamental approach can be based on the Beth–Uhlenbeck formula for the virial expansion.
Details can be found in the literature \cite{Roepke15}; see also \cite{SM}. % Appendix \ref{sec:BU}.
If one considers only the contribution from the bound states, the distribution function approaches zero smoothly at $P^{\rm Mott}_\nu$, rather than abruptly, when the Pauli blocking shift is taken into account.
As an example, we use the same parameter values $T=20$ MeV, $n_B=0.03$ fm$^{-3}$, $Y_p=0.5$ as in Fig. \ref{fig:dist-funct-wang}.
A calculation using the free values of the binding energy is compared with a calculation that takes the Pauli blocking shift according \cite{Roepke15} into account, see Fig. \ref{fig:distQS}.
The distribution is significantly reduced. The mass fraction $X_\alpha$ of the $\alpha$ particles decreases from 0.2212 to 0.01431. 
(The slight deviations in our calculation from the curve shown in Fig. \ref{fig:dist-funct-wang} are due to the fact that the DD2-RMF approach was used for the quasiparticle energies of the nucleons, rather than the Skyrme approximation. The Mott momenta of both approaches coincide for $\alpha$ particles, but differ slightly for $t$ and $d$.)

It is not only the $\alpha$-mass fraction $X_\alpha$ that decreases by a factor of approximately 15 when Pauli blocking is taken into account for the in-medium binding energies. 
Similarly, $X_t$ decreases from 0.0748 to 0.00323 and $X_d$ from 0.0883 to 0.000511.
As is known from the generalized Beth–Uhlenbeck formula, the dissolution of a bound state, when $P$ becomes smaller than the Mott momentum, leads to the formation of a resonance, such that the correlations of the continuum yield a finite value for the distribution function also beyond the Mott condition.
A consistent QS calculation of the composition of symmetric matter at various temperatures (including $T=20$ MeV) is shown in Fig. 4 of Ref.~\cite{Roepke15}.
As shown there, continuum correlations are relevant, particularly for the deuteron channel.

Both the Mott momentum and the distribution of light clusters determine the mass fraction $X_\nu$.
As an example, \cite{Wang2026} considers $npd$ matter; a comparison was carried out between the phase-space excluded-volume approach and an approach in which the in-medium energies $\epsilon_d$ are determined from the solution to the Schr\"odinger equation in the medium.
The good agreement shown there is possible if continuum contributions are neglected.
A well-studied example is the deuteron channel; see Fig. 7 in \cite{Schmidt90} for symmetric matter at $T=10$ MeV.
Above $n_B \approx 0.1\, n_{\rm sat}$, where deuterons with $P=0$ are resolved, contributions from finite total momentum are retained, and, in addition, contributions from scattering states become dominant, so that the maximum of the contribution from correlated states is reached at approximately $0.2\, n_{\rm sat}$.
The reference \cite{Wang2026} assumption of a negative contribution from scattering states is not supported by QS calculations. 
The contribution from $d$-like correlations becomes small at around $0.6\, n_{\rm sat}$, but never disappears entirely.
In \cite{Blaschke25}, the Mott lines were defined by the maximum of the fraction of correlated density at a given $T$.
A significant contribution from continuum correlations is related to resonances.
Well-known examples are $^4$H, $^5$He and $^8$Be, for which the contribution of continuum correlations can be calculated from the known scattering phase shifts \cite{Ropke20}.

The main discrepancies between the calculated compositions in the quantum statistical approach and in the phase-space excluded-volume approach (see Fig. \ref{fig:PLB25} and \ref{fig:PRL26}) can be attributed to the determination of the Mott momentum and the use of the energies $\epsilon_\nu(P)$ as its values in free space \cite{Wang2026}, whereby the Pauli blocking shift is neglected.
This leads to an overestimation of the mass fractions of light nuclei, in particular of $X_\alpha$,
whilst in \cite{Blaschke25} $X_\alpha$ becomes small when the baryon density exceeds the saturation density.
Furthermore, within the framework of an advanced approach, continuum correlations must be taken into account.

\begin{figure}[h]
  \centering
  \includegraphics[width=0.45\textwidth]{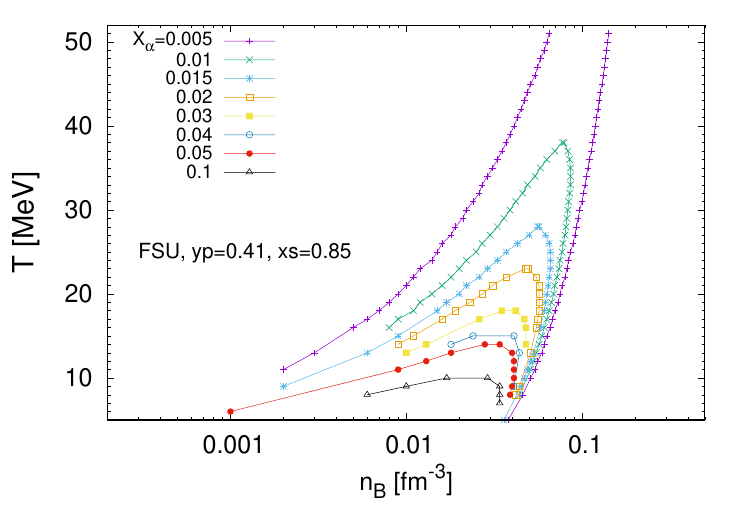}
\caption{Contour plots in the $(n_B,T)$ plane for different values of the mass fraction of the $^4$He cluster within the gRMF approach, fixing the proton fraction to 0.41, and taking the FSU interaction. The scalar cluster-meson coupling is equal to 0.85.}
\label{fig:T_contour}
\end{figure}

As an alternative to the phase-space excluded volume approach, we discuss another simple method for determining the composition of the nuclear matter: the generalised RMF approach. 
Details are given in \cite{Pais2018}, see \cite{SM}. % Appendix \ref{app:gRMF}.
For the light clusters, quasi-particle energies are assumed. The dissolution of the clusters is affected
by a combination of the scalar cluster-meson coupling factor and the binding energy shift, a term that acts as the energetic counterpart of the excluded volume mechanism in the Thomas-Fermi approximation.
%The shifts determine the Mott density. 
A contribution to the composition can also be made beyond the Mott density, thereby simulating the contribution of continuum correlations.

\begin{figure}[h]
  \centering
  \includegraphics[width=0.5\textwidth]{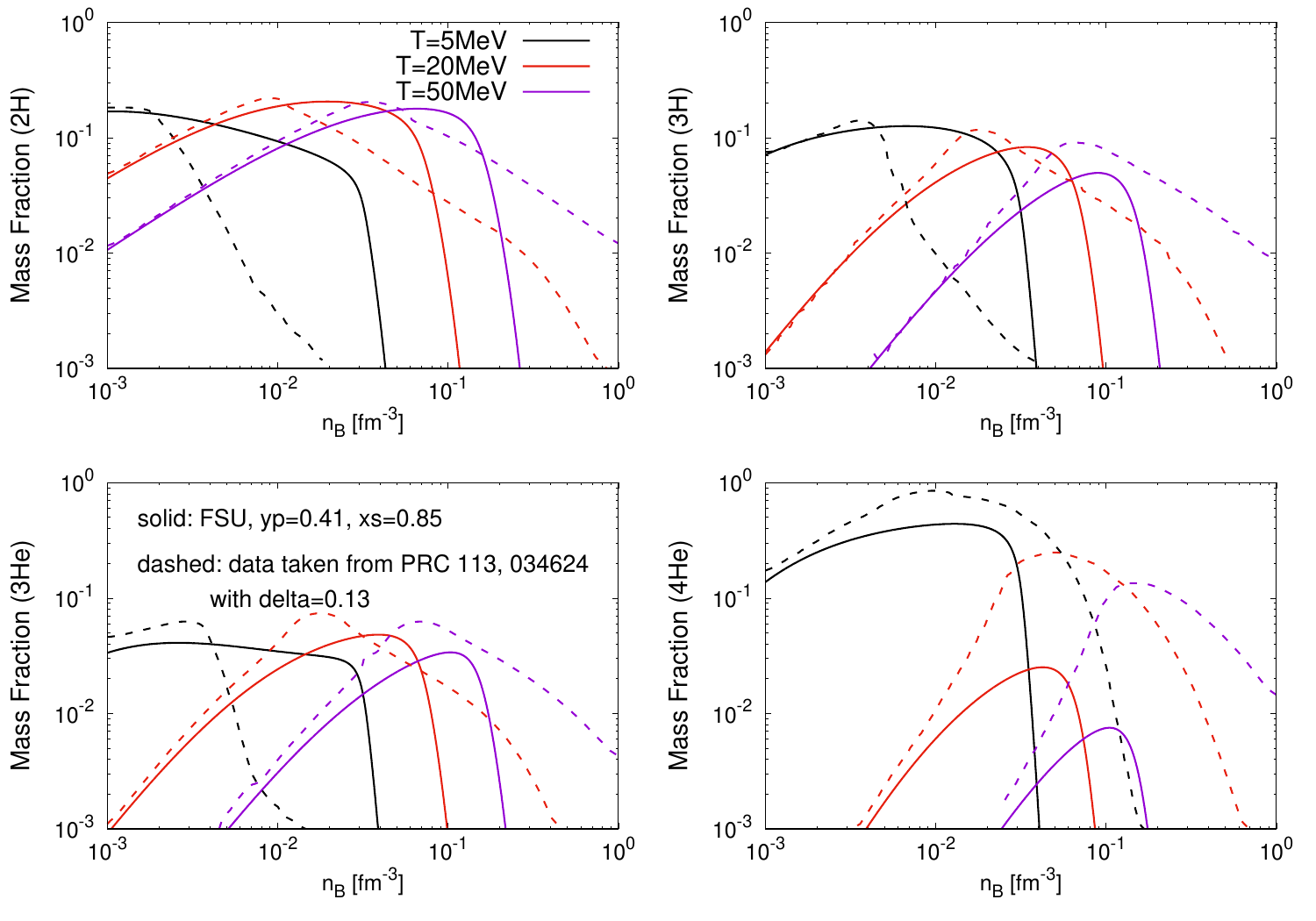}
\caption{Mass fractions of the four light clusters as  a function of the baryonic density for $T=5$ (black), 20 (red) and 50 (blue) MeV. The solid lines are from the gRMF calculation, taking FSU as the nuclear interaction, the proton fraction is fixed to 0.41, and the scalar cluster-meson coupling is equal to 0.85. The dashed lines are results for the PSEV approach, taken from Ref.~\cite{Wang2026}, with the isospin asymmetry $\delta=0.13$.}
\label{fig:mass_fractions-comparison}
\end{figure}

We have also performed a calculation of the $\alpha$ mass fractions $X_\alpha$ within the gRMF approach, see Fig.~\ref{fig:T_contour}.
Comparing with Fig.~\ref{fig:PLB25}, a reasonable agreement with the QS approach is found. However, looking at Fig.~\ref{fig:PRL26}, a large discrepancy with these calculations  is observed.
A comparison of the mass fractions as a function of density for given $T, Y_p$ is shown in Fig.~\ref{fig:mass_fractions-comparison} %\ref{fig:mass_fractions-prc-wang}, %\ref{fig:mass_fractions}
for the phase-space excluded-volume approach \cite{Wang2026} (dashed) and the gRMF calculations (solid). Although the proton fraction $Y_p$ in these two calculations, as well as the interactions (Skyrme vs. RMF) taken into account, differs only slightly, there is a rather large difference in composition, particularly for $X_\alpha$.

When comparing different approaches to describing hot and dense nuclear matter, we conclude that the Pauli blocking shift in the in-medium binding energies of light clusters must be taken into account.
Whilst such quasiparticle shifts in the binding energies of light clusters are described in the gRMF approximation, this is absent within the present PSEV calculations \cite{Wang26,Wang2026}.
These significant differences are also reflected in the contour plots (Figs. \ref{fig:PLB25} – \ref{fig:PRL26}) shown at the beginning of the article.
On the other hand, the phase-space excluded-volume is capable of describing the pronounced momentum dependence of the Pauli blocking shift.
Further improvements are needed to reproduce the results of a systematic quantum-statistical approach.

\section{Discussion: Extracting astrophysically useful information from HIC experiments}

We have discussed the equilibrium properties—in particular, the composition—of hot and dense nuclear matter.
The results are of interest for astrophysical applications like binary neutron star mergers \cite{Sneppen:2024jch}.
In laboratory experiments, hot and dense matter is produced during heavy-ion collisions.
However, the state of the matter is inhomogeneous in both space and time, necessitating a non-equilibrium approach.
Local thermodynamic equilibrium properties are merely a prerequisite for describing the non-equilibrium evolution of matter in heavy-ion collisions.

There are numerous codes which describe the evolution of hot and dense matter produced in HIC as reaction-kinetic equations, such  as %e.g., 
the generalized BUU equations, the UrQMD+GEMINI$^{++}$ code (see \cite{Xian23} and further references given there), or SMASH (see \cite{Sasha26} and further references given there). %, etc., 
These approaches give also the composition of hot and dense matter, but the systematic treatment of correlations in a hot and dense medium remains an open problem.
However, in heavy-ion collisions which are non-equilibrium processes, reaching equilibrium properties is a necessary benchmark. 

For expanding hot and dense matter, the freeze-out concept was employed.
Local thermodynamic equilibrium is nearly achieved when the intrinsic relaxation time is shorter than the relative change in the thermodynamic parameters.
After freeze-out, a kinetic description is possible that takes into account the distribution function of the constituents. In particular, the decay of excited states takes place.

The phase-space excluded-volume approach \cite{Wang2026}, in which a cutoff value $F^{\rm cut}_A$ is introduced, can be viewed as a simple method for reproducing the effects of Pauli blocking.
It was used to formulate a kinetic approach \cite{Wang26,Wang2026},
which was employed to discuss the FOPI experiments \cite{Reisdorf10}.
This approach could be promising, but it requires further improvements, such as fully accounting for Pauli blocking and incorporating excited states as well as larger clusters.

To explain the lower energy NIMROD data \cite{Qin12,Hagel12}, the momentum dependence of the binding energy shifts, the contribution of continuum correlations, and the correlations within the medium were taken into account. 
A similar analysis has been performed with INDRA data \cite{Bougault,PaisPRL}.
Both experiments probe temperatures below 10 MeV. It would be of interest to extend Figs. 2  and 3 of  \cite{Wang26} to the interval $0 \le T \le 20$ MeV.
In a more recent analysis \cite{Tiago1,Tiago2} of the central Xe+Sn collisions from INDRA data, 
 an excellent reproduction of the experimentally measured abundances of the H and He isotopes has been obtained, and a faster decrease of the light clusters abundances with temperature has been observed in contrast to \cite{Wang26}.

Although the high experimentally derived $\alpha$ particle fraction is attributed to high densities,  
as shown by the confidence band  in Fig. 3 of  \cite{Wang26}, determining its origin requires further information.
A crucial point is the determination of $n_B$ from experimental yields.
Furthermore, a significant portion of the apparent “excess” of $\alpha$ particles could be attributable to the feed-down caused by the decay of excited $A=5-8$ fragments, see  \cite{Natowitz23} for $T \approx 1$ MeV. For $T$ in the range of 20 - 40 MeV, more excited states of the nuclei must be taken into account.
As mentioned in \cite{Wang26}, short-range correlations \cite{Hen17,Burrello22} should also be considered to calculate in-medium effects. 
An earlier analysis of the FOPI experiments using AMD \cite{Ono17,Ono19,Cheng24} confirmed the in-medium modifications of the properties of light clusters.

In conclusion, more fundamental investigations are required before abundant $\alpha$ clustering above the saturation density is confirmed.
In this paper, we have focused on the properties of local equilibrium. A discussion of the HIC experiments—in particular, the extraction of information useful for astrophysics—will be the subject of future work.

\section*{Acknowledgements}

We thank R. Wang and his co-authors for very useful discussions and clarifications during the preparation of this paper. 
This work was partially supported by Portuguese national funds from FCT (Fundação para a Ciência e a Tecnologia, I.P., Portugal) under project 2024.16290.PEX with DOI identifier 10.54499/2024.16290.PEX and under project UID/04564/2025, identified by DOI 10.54499/UIDB/04564/2025. 
G.~R.~acknowledges a honorary stipend from the Foundation for Polish Science within the Alexander von Humboldt program under grant No. DPN/JJL/402-4773/2022.

\section*{Data availability statement}
Further data can be found in the supplementary material accompanying this article \cite{SM}. Additional data relating to this work are available from the authors upon reasonable request.

\end{document}